\documentclass[preprint2,twoside]{hwo}

\usepackage{lipsum}

\input{hwo.h}

\begin{document}

\title{\textbf{\LARGE Prebiosignatures with the Habitable Worlds Observatory}}
\author {\textbf{\large Sukrit Ranjan$^{1,2}$, Danica Adams$^3$, Michael Wong$^4$, Martin Schlecker$^{5,6}$, Nicholas Wogan$^7$, Jessica M Weber$^8$}}
\affil{$^1$\small\it Lunar \& Planetary Laboratory, University of Arizona, Tucson, AZ, USA}
\affil{$^2$\small\it Blue Marble Space Institute of Science, Seattle, WA, USA}
\affil{$^3$\small\it Dept. of Earth \& Planetary Sciences, Harvard University, Cambridge, MA, USA}
\affil{$^4$\small\it Earth \& Planets Laboratory, Carnegie Institution for Science, Washington, DC, USA}
\affil{$^5$\small\it Steward Observatory, University of Arizona, Tucson, AZ, USA}
\affil{$^6$\small\it European Southern Observatory, Karl-Schwarzschild-Strasse 2, Garching by Munich, Germany}
\affil{$^7$\small\it Space Science Division, NASA Ames Research Center, Moffett Field, CA, USA}
\affil{$^8$\small\it  NASA Jet Propulsion Laboratory, California Institute of Technology, Pasadena, CA, USA}



\author{\footnotesize{\bf Endorsed by:}
Steven Dillmann (Stanford University), Kaustubh Hakim (KU Leuven/Royal Observatory of Belgium), Chris Impey (University of Arizona), Gaetano Sandariato (INAF), Eunjeong Lee (EisKosmos (CROASAEN), Inc.), Farod Salama (NASA Ames Research Center), Celia Blanco (BMSIS), Blair Russell (Chapman University), Oliver Carey (Brown University), Melinda Soares-Furtado (UW-Madison), Iva Vilovi\'c (Leibniz Institute for Astrophysics Potsdam), Finnegan Keller (Arizona State University), Aarynn Carter (STScI), Jens Kammerer (European Southern Observatory), Elena Manjavacas (STScI), Ana Ines Gomez de Castro (Universidad Complutense de Madrid), Faraz Nasir Saleem (Egyptian Space Agency), William Roberson (New Mexico State University), Theresa Fisher (University of Arizona), Katherine Bennett (Johns Hopkins University), Natalie Allen (Johns Hopkins University), Xinchuan Huang (SETI Institute \& NASA Ames Research Center), Joshua Krissansen-Totton (University of Washington), Tomas Stolker (Leiden University), Austin Ware (Arizona State University), Arnaud Salvador (German Aerospace Center).
}

\begin{abstract}
The Habitable Worlds Observatory (HWO) aims to characterize habitable exoplanets in search of signs of life. However, detectable life may be rare, either because abiogenesis is intrinsically contingent and unlikely, or because biospheres may efficiently recycle their products. Here, we explore the potential of HWO to test theories of life in the universe even if detectable life is rare by searching for “prebiosignature gases”. Prebiosignatures gases are gases whose detection constrains theories of the evolution of prebiotic (habitable but uninhabited) planets, thereby testing theories of abiogenesis and guiding laboratory investigations of the origin of life. We catalog 5 theories of prebiotic environments that are potentially testable by HWO, identify their observational tests, and rank them by perceived detection plausibility. The prebiosignature paradigm is novel and potentially compelling, but considerable work is required to mature it and assess its practical relevance for HWO, especially simulated spectral observation and retrieval studies. However, consideration of the absorption properties of prebiosignature observables alone reveals that coverage at NUV wavelengths (200-400 nm) will be required to effectively realize a prebiosignature science case for HWO, supporting the argument for UV capabilities for HWO.   \\
  \\
  \\
\end{abstract}

\vspace{2cm}

\section{Science Goal:  \textmd{Test theories of prebiotic planetary conditions.}}
The search for life on exoplanets with telescopes is subject to an inherent risk, which is the possibly contingent\footnote{i.e. subject to chance.} nature of the emergence and evolution of life. It is possible that the emergence of life is an intrinsically unlikely event, and that therefore no planet in our solar neighborhood hosts biology (e.g., \citealt{Rimmer2023}). Even if the emergence of life is common, it is possible that the evolutionary histories of exobiology result in biospheres in which the products of biology are efficiently recycled without affecting the planetary envelope, resulting in ``false negatives" for life search \citep{Reinhard2017}. However, even in these cases HWO may constrain theories of life in the universe by detecting testing theories of prebiotic environments, guiding laboratory studies of prebiotic chemistry (e.g., \citealt{Claringbold2023, Keller2025}). Synthetic spectra and retrieval analysis are required to substantiate this possibility.

\section{Science Objective: \textmd{Measure the atmospheric composition of prebiotic exoplanets in search of prebiosignature gases}\label{sec:sciobj}}

\begin{figure*}[ht]
\begin{center}
\includegraphics[width=0.6\textwidth]{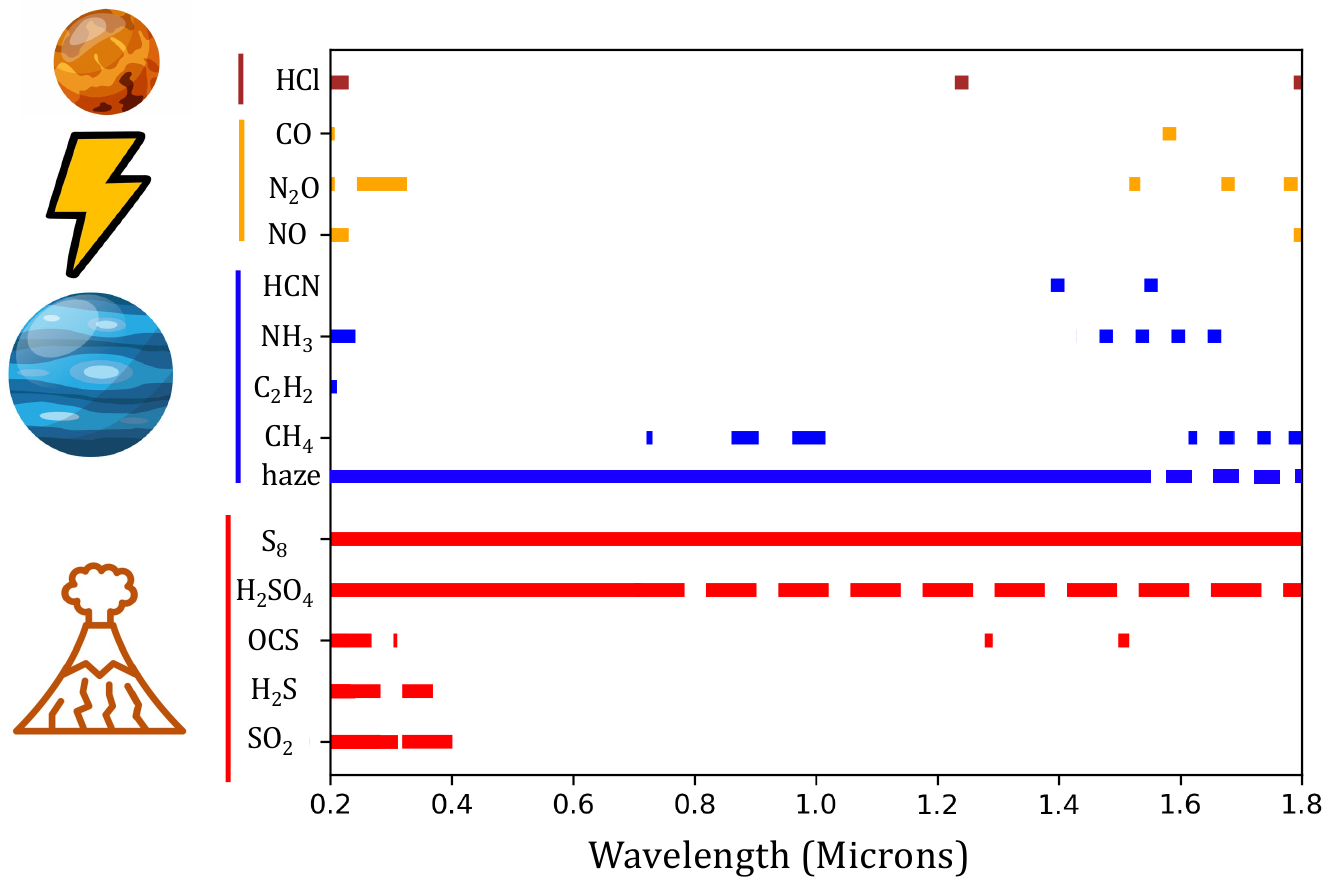}
\caption{\small Prebiosignature atmospheric species and the wavelengths where they absorb or scatter (extinct) light. Wavelengths where species extinction exceeds $10^{-24}$ cm$^{-2}$ molecule$^{-1}$ and $10^{-18}$ cm$^{-2}$ molecule$^{-1}$ are indicated by dashed lines and solid lines, respectively. Species are grouped by the origin-of-life scenario they correspond to. The UV bandpass will be key to detecting prebiosignature gas species. Underlying extinction data are from \citet{Arney2016, Hu2013, Ranjan2017a, Keller-Rudek2013}. Underlying extinction data are limited to what is available in the literature; it is possible that these species absorb at other wavelengths as well, but such absorption is not yet known.  Figure motivated by Figure 2 of \citet{Rimmer2021}. \label{fig:prebiosig1}
}
\end{center}
\end{figure*}

HWO may potentially constrain theories of the origins of life by searching for ``prebiosignature gases", i.e. gases which constrain theories of the atmospheric evolution of prebiotic planets (Figure~\ref{fig:prebiosig1}). By ``prebiotic planet", we refer to planets which are habitable but not yet inhabited. Early Earth prior to the origin of life was a prebiotic planet, and prebiotic Earth's atmosphere is extensively invoked as sources of molecular reagents to initiate the origin of life based on model calculations (e.g., \citealt{Miller1953, Pinto1980, Mancinelli1988, Ranjan2023c}). However, there is considerable uncertainty and even disagreement in these model calculations due to imperfect understanding of the relevant physical and chemical processes (e.g., \citealt{Ranjan2023c}), and direct geological tests are difficult due to extensive processing of the rock record by plate tectonics and hydrology \citep{Sasselov2020}. HWO observations of habitable planet atmospheres offer the possibility of testing these model predictions \citep{Keller2025, GoodisGordon2025}.

\section{Physical Parameters}
\subsection{Summary}
\begin{figure*}[ht]
\begin{center}
\includegraphics[width=0.6\textwidth]{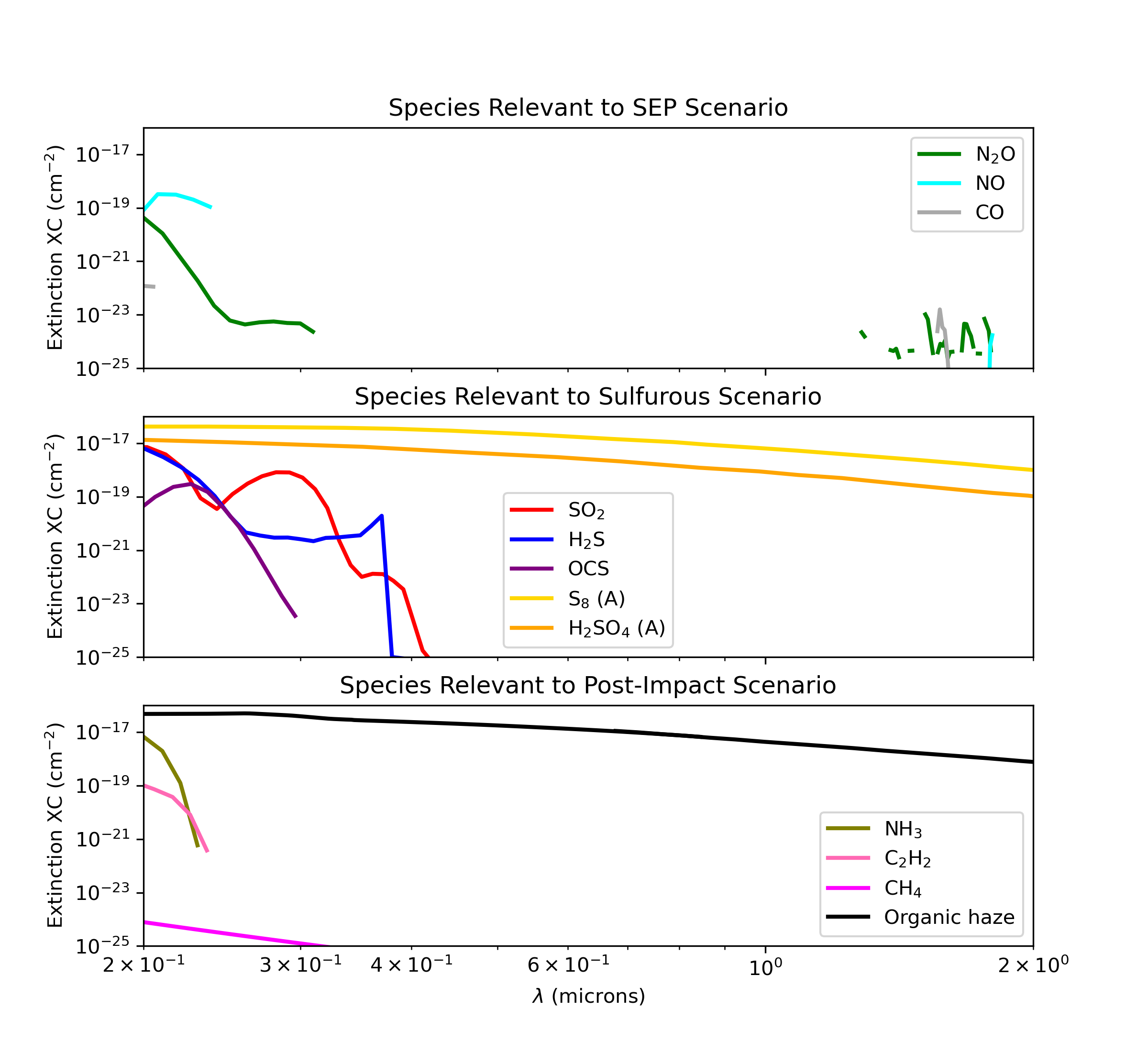}
\caption{\small Prebiosignature atmospheric species and their extinction as a function of wavelength. Species are grouped by the origin-of-life scenario they correspond to. Underlying extinction data are from \citet{Arney2016, Hu2013, Ranjan2017a, Keller-Rudek2013}. Underlying extinction data are limited to what is available in the literature; it is possible that these species absorb at other wavelengths as well, but such absorption is not yet known. Simulated spectral and retrieval analysis is required to determine the feasibility of observing these
molecules with HWO.
\label{fig:prebiosig3}
}
\end{center}
\end{figure*}

\begin{itemize}
\item Obtain atmospheric spectra of prebiotic planets down to $0.2~\mu$m. Search these spectra for constraints on prebiosignature gases (Figures
~\ref{fig:prebiosig1},~\ref{fig:prebiosig2}). This science case will be more challenging with a cutoff of $0.25~\mu$m due to weaker absorption of all prebiosignature gases (Figure~\ref{fig:prebiosig3}), and the ability to characterize NO, NH$_3$, and C$_2$H$_2$ will be ceded altogether.
\end{itemize}
As stated in Section~\ref{sec:sciobj}, we define ``prebiotic" planets as those which are habitable but uninhabited\footnote{While more environmental conditions than just habitability may be required for abiogenesis \citep{Deamer2022, Wong2022, Keller2025}, the additional conditions are highly specific to proposed origin-of-life hypotheses and are often contradictory. We therefore argue it is more productive to explore how HWO observations can reveal the requirements for abiogenesis; see the ``Testing Origin-of-Life Theories with HWO" Science Case Development Document.}.  Observationally, we suggest planets should be considered candidate ``prebiotic" planets if they show evidence for habitable conditions (e.g. presence of surface liquid water) but lack evidence for biology (e.g., lack of biosignatures). Younger planets ($<1$ Gyr) are also more likely to be prebiotic, since finite time is required for life to emerge

\subsection{Extended}
\begin{figure*}[ht]
\begin{center}
\includegraphics[width=0.6\textwidth]{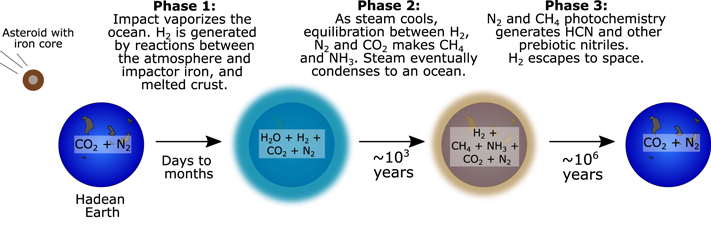}
\caption{\small Schematic of the post-impact scenario for prebiotic atmospheres \citep{Genda2017, Zahnle2020, Wogan2023}. In this scenario, reducing power is delivered to the surface-atmosphere system by reaction of an iron-rich meteor with the planetary ocean, generating a transient Miller-Urey type atmosphere. We anticipate such atmospheres to be strongly distinguishable by HWO observations through their copious concentrations of reducing species \citep{GoodisGordon2025}. However, we also anticipate such atmospheres to be rare due to their transient nature, and a sample of 25 planets is unlikely to capture them. Figure taken from \citet{Wogan2023}.
\label{fig:prebiosig2}
}
\end{center}
\end{figure*}

Here, we discuss different theories of prebiotic environments that HWO observations could test. There is a trade-off: the prebiosignatures that we expect to be more readily detectable are also expected to be rarer, and may not be accessible with a nominal 25-planet sample. We present the prebiosignature target possibilities below, ordered by perceived frequency. We emphasize that this ordering is mostly based on qualitative intuition, not quantitative analysis, and additional work is required to rigorously constrain it.
\begin{itemize}
\item \textbf{Steady-State Prebiotic Atmospheres Are be H$_2$-Poor and CO$_2$/N$_2$ Rich:} In the steady-state, abiotic habitable planet atmospheres are predicted to be H$_2$-poor ($<1\%$ H$_2$) and primarily composed of CO$_2$ and N$_2$ \citep{Gaillard2014, CatlingKasting2017, Liggins2020, Hu2023}. This leads to the prediction that rocky planets should
generally host H$_2$-poor atmospheres, with the main potential exception being in the transient aftermath of large impacts (see below). HST and JWST have already begun to confirm this prediction for low-mass stars \citep{deWit2018, Hu2024}; HWO can extend this test to Solar-type stars. Ubiquity of CO$_2$ and N$_2$ on temperate terrestrial planets and nondetections of H$_2$ would be consistent with this hypothesis, while detection of abundant H$_2$ on temperate terrestrial exoplanets, e.g. via the 1.2 $\mu$m H$_2$-H$_2$ CIA feature \citep{Borysow2002, Koroleva2024} would falsify this theory.
\item \textbf{Steady-State Prebiotic Atmospheres Are Nonreducing:}  In the steady-state, abiotic habitable planet atmospheres are generally predicted to be poor in HCN and CH$_4$
\citep{Zahnle1986, Guzman-Marmolejo2013, Rimmer2019, Pearce2022}. This is important because these gasses are heavily invoked in prebiotic chemistry. Searches for HCN and CH$_4$ on temperate terrestrial planets can test this prediction.
\item \textbf{Steady-State Prebiotic Atmospheres Should Be Sulfur-Poor:} There is uncertainty about sulfur cycling on abiotic planets, with the main uncertainty being whether habitable planets (temperate planets with oceans) can accumulate sulfur in the atmosphere in the form of SO$_2$ or sulfur haze (H$_2$SO$_4$, S$_8$), or whether the ocean efficiently absorbs sulfur and prevents its accumulation \citep{Kasting1989, Kaltenegger2010, Hu2013, Ranjan2023c, Loftus2019}. This is relevant to prebiotic chemistry due to controls on UV light and sulfur availability for prebiotic chemistry \citep{Ranjan2018}. The uncertainty is driven by severe disagreement in the kinetics of oceanic chemistry, with the latest entry in the debate arguing the ocean should absorb the atmospheric sulfur \citep{Ranjan2023c}. HWO can test this theory by testing whether SO$_2$ and/or H$_2$SO$_4$/S$_8$ aerosols are anti-correlated with
liquid water ocean indicators, as the theory predicts, and more generally whether atmospheric sulfur is common on temperate terrestrial planets.
\item \textbf{Prebiotic Atmospheres Fix Nitrogen in the Form of Oxidized Nitrogen Species:} The atmosphere is generally invoked as the source of prebiotic fixed nitrogen for nascent life, through the mechanisms of lightning fixation and Solar Energetic Particle (SEP) fixation \citep{Mancinelli1988, Airapetian2016}. For example, lightning is proposed to
recombine CO$_2$ and N$_2$ into NO and CO \citep{Mancinelli1988, Harman2018}. This mechanism may be detectable through observation of NO in exoplanet atmospheres, though the $\sim10^{8}\times$ uncertainty in lightning-derived NO abundance must first be reconciled \citep{Ardaseva2017, Wong2017, Barth2024}. Indeed, observations of NO have been invoked as evidence of nitrogen fixation on Venus, through either lightning or impact synthesis \citep{Krasnopolsky2006, Blaske2023}. Similarly, SEPs are predicted to fix nitrogen in the early Earth's atmosphere, up to 1 ppm N$_2$O (up to 1000 ppm in the stratosphere) \citep{Airapetian2016, Airapetian2020, Schwieterman2022}. Detection of N$_2$O in the atmospheres of planets orbiting active stars and its nondetection in planets orbiting inactive stars may test this paradigm
also.
\item \textbf{Prebiotic Atmospheres Are Transiently Reducing in the Aftermath of Large Impacts:} Atmospheric synthesis after impact of an iron-rich asteroid is often invoked as
a mechanism for generation of the highly reduced compounds generally demanded by prebiotic chemistry \citep{Genda2017, Zahnle2020, Wogan2023}. The basic proposed mechanism is the liberation of reducing power in the form of H$_2$ by reaction of the water envelope with the metallic iron from the impactor (Figure~\ref{fig:prebiosig2}). However, these calculations assume that the iron delivered by the impactor reacts to completion with the oceanic envelope. This need not be the case; for example, the impact may deposit the iron into the mantle of the planet directly \citep{Citron2022, Itcovitz2022}, or the iron may develop a crust which prevents the reaction from going to completion before it is subducted away from the surface. HWO can search for evidence of such transient post-impact atmospheres by looking for H$_2$-dominated atmospheres with large amounts of methane (CH$_4$; $\sim 0.001$ v/v, \citealt{Wogan2023}), hydrogen cyanide (HCN), acetylene (C$_2$H$_2$), cyanoacetylene (HC$_3$N), ammonia (NH$_3$), and organic haze, which are otherwise predicted to be rare \citep{Kasting1982, Guzman-Marmolejo2013, Rimmer2023, Rimmer2021, Pearce2022, Claringbold2023, Wogan2023}. While readily identifiable spectroscopically \citep{GoodisGordon2025}, such atmospheres are transient and geologically short-lived. Earth spent a cumulative 0.01-0.1 Gyr in a reducing H$_2$-dominated atmosphere on Earth \citep{Zahnle2020}, with the impacts confined to the first 0.6 Gyr of Earth's history to 95\% confidence \citep{Wogan2024}. Assuming a random distribution of planet age from 0-10 Ga, this implies that 100-1000 planets must be observed to image one in a post-impact atmosphere state, much higher than the nominal HWO goal of characterizing 25 planets. The best chance for observing post-impact atmospheres would entail systematically observing planets in young systems ($<1$ Gyr), for which 10-100 planets would need to be imaged. While such observations entail getting lucky, they are not impossible; for example, the transient signature of a giant impact during planet formation was recently observed with ALMA \citep{Schneiderman2021}. 
\end{itemize}

\section{Description of Observations}
\subsection{Summary}

\subsubsection{HWO Observations}

\begin{itemize}
\item High-precision UV-VIS spectra of temperate terrestrial planets. Note that the prebiosignatures absorb most strongly in the UV part of the spectrum, making UV
capability down to 200 nm critical for achieving this science case (Figures~\ref{fig:prebiosig1}, \ref{fig:prebiosig2}).
\end{itemize}

\subsubsection{Useful Complementary Observational Data}
\begin{itemize}
\item Ages of stellar systems (young systems more likely to be prebiotic, have high impact and volcanism rates). Specifically, discriminating stars with age $>1$ Gyr
from stars with age $<1$ Gyr, motivated by the early emergence of life on Earth ($<1$ Gyr age; \citealt{Ranjan2017a, Pearce2018}).
\item Confirm presence of habitable conditions on planet (e.g., via glint or characterization of vertically-resolved H$_2$O profiles; \citealt{Robinson2018})
\item Lack of biosignatures (surface, atmospheric).
\end{itemize}

\subsubsection{Suggestions for Future Work}
\begin{itemize}
\item Need to generate synthetic spectra, run retrieval analysis on prebiotic scenarios.
\end{itemize}

\subsection{Extended}
A key limitation of the prebiosignature science case is the lack of study of observability of prebiosignatures (synthetic spectra, retrievals). Such studies are lacking because the community incepting origin-of-life hypotheses has been focused on early Earth, for which remote observation is not relevant. Studies to determine the spectral features of the gases and hazes described above and their detectability at plausible concentrations are required, as are standardly conducted for biosignature gases and habitability indicators. The first such studies have been conducted, focused on JWST observables of prebiosignatures (e.g., \citealt{Rimmer2023, Claringbold2023}); more studies focused on the UV-VIS-NIR HWO bandpass are required.

\subsection{Impact on Requirements}
Table~\ref{tbl:req} provides the impacts of this science case on the requirements, following the HWO SSSCDD template. 

\begin{table*}[!ht]
\caption{Impact on Requirements\label{tbl:req}}
\smallskip
\begin{center}
{\small
\begin{tabular}{|p{2.2cm}|p{1.5cm}|p{1.5cm}|p{7.0cm}|p{2.0cm}|}  
\tableline
\textbf{Capability or Requirement} & \textbf{Necessary} & \textbf{Desired} & \textbf{Justification} & \textbf{Comments}\\
\tableline
UV Observations & Yes & -- & Most known absorption for prebiosignature gases is in the UV & Figures~\ref{fig:prebiosig1}, \ref{fig:prebiosig2}\\
\tableline
Long wavelength observations ($>1.5~\mu$m) & -- & Yes & Some known absorption for prebiosignature gases is in the NIR & Figures~\ref{fig:prebiosig1}, \ref{fig:prebiosig2}\\
\tableline
Timing of observations in different bands &  No & No  & We do not expect significant changes in bulk atmospheric composition on the timescale of HWO lifetime & \\
\tableline
High Spatial Resolution &  No & No  & Disk-integrated observations are assumed for this science case & \\
\tableline
High Spectral Resolution &  Unknown & -- & Simulated spectra/retrievals required to quantify necessary spectral resolution & \\
\tableline
Large Field of View &  No &  No & N/A & \\
\tableline
Rapid Response &  No &  No & We do not expect significant changes in bulk atmospheric composition on the timescale of HWO lifetime & \\

\tableline
\end{tabular}
}
\end{center}
\end{table*}

%




{\bf Acknowledgements.} SR gratefully thanks the University of Arizona for support via startup. JMW carried out this work at the Jet Propulsion Laboratory, California Institute of Technology, under a contract with NASA (80NM0018D0004). D.A.’s and MLW’s research is funded by NASA through the NASA Hubble Fellowship Program Grants HST-HF2-51523.001-A and HST-HF2-51521.001-A, respectively, awarded by the Space Telescope Science Institute, which is operated by the Association of Universities for Research in Astronomy, Inc., for NASA, under contract NAS5‐26555.

\bibliography{prebiosignatures.bib}

\end{document}